# Decadal wave reconstruction in the Mediterranean Sea with graph neural networks

Federica Benassi,[a,b] Lorenzo Mentaschi,[b,a] Salvatore Causio,[a] Daniel Holmberg,[c] Ivan Federico,[a] Nadia Pinardi[b,a].

[a] *CMCC Foundation – Euro-Mediterranean Center on Climate Change, Lecce, Italy*

[b] *Department of Physics and Astronomy, University of Bologna, Bologna, Italy*

[c] *Department of Computer Science, University of Helsinki, Helsinki, Finland*

*Corresponding author*: Federica Benassi, federica.benassi@cmcc.it

ABSTRACT

Accurate simulation and prediction of ocean waves are essential for coastal risk management and climate studies. Deep learning has shown promising results for wave modeling, but most approaches still operate on regular grids and on forecasting time scales, and do not generalize to unstructured discretization or to long time horizons. Here we present WaveGraph, a model based on Graph Neural Networks (GNNs) that emulates basin-scale wave dynamics directly on unstructured meshes with high resolution along the coasts (up to 2-3 km). Trained on bias-corrected simulation data over the Mediterranean Sea, WaveGraph uses a multiscale architecture combining the unstructured model mesh with a uniform graph, allowing simultaneous representation of local coastal interactions and large-scale wave dynamics. The model reconstructs the evolution of significant wave height, mean period, and mean direction, and is applied autoregressively for a continuous 17-year period without reinitialization or drift. Validation against buoy and satellite observations shows skill comparable to the input data set, and ablation experiments indicate that wind forcing drives most of the long-term stability while wave history improves swell-driven and basin-scale dynamics. These results show that GNNs can provide stable and efficient emulators of spectral wave models on unstructured domains, enabling decadal wave reconstructions.

*This* Work *has been submitted to Artificial Intelligence for the Earth Systems.*

## 1. Introduction

Ocean surface waves play an important role in the Earth system by mediating the exchange of momentum and energy between the atmosphere and the ocean, influencing weather and climate dynamics, and shaping coastal processes. They affect offshore operations and navigation, drive sediment transport and shoreline evolution, and contribute to coastal flooding (Causio et al., 2025; Cavaleri et al., 2007; Holthuijsen, 2007; Komen et al., 1994; Lopes et al., 2026; Shirinov et al., 2025). Furthermore, waves are key elements of air-sea exchanges of momentum, heat and chemical tracers, driving the ocean circulation and being part of modern operational forecasting models (Clementi et al., 2017; Causio et al., 2021, Coppini et al., 2023). Wave conditions are inherently non-stationary, responding to both natural climate variability and anthropogenic climate change, with observed trends and projected future changes varying across ocean basins (Casas-Prat et al., 2024; Vousdoukas et al., 2018; Young et al., 2011). For these reasons, accurate simulation and prediction of ocean waves are essential for operational forecasting, coastal hazard assessment, and understanding the evolving climate system.

State-of-the-art wave prediction systems are based on numerical solutions of the phase-averaged spectral action density balance equation, which describes the evolution of wave energy under wind forcing, nonlinear wave-wave interactions, and dissipation processes (Holthuijsen, 2007). These spectral models have achieved high predictive skill and are widely used for both operational and research applications (Booij et al., 1999; Komen et al., 1994; Tolman et al., 2002). Recent advances have extended these models to unstructured meshes, allowing high spatial resolution in coastal regions while maintaining coarser resolution offshore and improving the representation of complex coastlines and bathymetry (Zijlema, 2010; Roland et al., 2012; Lira-Loarca et al., 2022; Toomey et al., 2022; Mentaschi et al., 2023). However, the computational cost of high-resolution, basin-scale unstructured grid wave simulations and ensemble experiments remains a major limitation. For this reason, wind-wave processes have generally been excluded from coupled Earth system models, despite their recognized role in air-sea fluxes and coupled climate dynamics (Brus et al., 2021).

Data-driven approaches have emerged as alternatives or complements to traditional numerical modeling in atmospheric and ocean sciences. Machine learning models can leverage large observational and numerical model datasets predicting wave conditions at low computational cost while remaining flexible to different types of data. Early applications in wave modeling focused on point-based predictions, where neural networks estimate wave parameters at individual locations using local wind conditions or nearby observations (Berbić et al., 2017; Jörges et al., 2021). More recent work has extended these approaches to spatial domains on structured grids using multilayer perceptrons (e.g. James et al., 2018), convolutional neural networks (Zhou et al., 2021; Ouyang et al., 2023; Cao et al, 2023; Zhang et al., 2025; Wang et al., 2025), and by incorporating wave forecasting into Earth system foundation models (Bodnar et al., 2025). These studies have also shown that machine learning models can be used in autoregressive settings to produce predictions, with recent work demonstrating stable predictions over time horizons exceeding 10 days (Wang et al., 2025). However, these approaches have been developed for wave prediction on regular rather than unstructured meshes.

Graph neural networks (GNNs) provide a natural framework for learning spatiotemporal dynamics and have been successfully applied to several Earth system components, including the atmosphere (Keisler, 2022; Lam et al., 2023; Oskarsson et al., 2024) and the ocean (Hahner et al., 2026; Holmberg et al., 2025, 2026). GNNs operate directly on graph representations of the domain, where nodes correspond to spatial locations and edges define their connectivity. This makes them particularly well suited for unstructured meshes, where spatial relationships are irregular and resolution varies across the domain.

A limited number of studies have applied GNNs to wave modeling on unstructured meshes. These works have primarily focused on spatial reconstruction or limited-area downscaling (Kuehn et al., 2024), or on relatively small, fetch-dominated domains (Iqra et al., 2025), rather than learning the full spatiotemporal evolution of wave fields at basin scales. Additional studies have considered graph representations based on observational buoy networks (Lin et al., 2025; Yin et al., 2024; Zhang et al., 2023), which do not resolve the full spatial structure of the wave field.

In this study, we develop WaveGraph, a GNN designed to emulate the dynamics of a phase-averaged spectral wave model on an unstructured mesh on the entire Mediterranean Sea. The model represents the computational mesh as a graph that combines fine coastal resolution with a coarser basin-scale structure, allowing the representation of both local and large-scale wave interactions. It is trained on the Wadalkar et al. (2026) dataset, a bias-corrected version of the Mentaschi et al. (2023) simulation, to predict the one-step evolution of multiple wave parameters, including significant wave height, mean wave period, and mean wave direction, and is applied autoregressively to generate long-term, decadal reconstructions of wave conditions. The model is trained and validated on a period of 6 years and the autoregressive reconstruction is shown for the following 17 years (2007-2023). Ablation experiments show the sensitivity of WaveGraph to different input fields.

The paper is organized as follows. Section 2 describes the WaveGraph architecture, training data, and evaluation metrics. Section 3 presents the main results, including performance against the reference model, buoy and satellite validation, and an ablation experiment isolating the contribution of wind forcing. Section 4 discusses the main drivers of model stability and limitations. Section 5 summarizes the conclusions and outlines directions for future work.

## 2. Methods

WaveGraph is designed as a data-driven emulator of spectral wave dynamics on an unstructured spatial mesh. At each mesh node, the wave state $X^t$encodes the predicted variables at time $t$; the temporal evolution of this state is driven by atmospheric forcing $F^t$ (here, the 10 m wind field) and by static spatial features $c$ such as bathymetry and node coordinates.

Rather than learning the full state directly, the model approximates a one-step transition function $f$ that maps the current wave state and a short window of atmospheric forcing to the next state:

$$\Delta X^{t+1} = X^{t+1} - X^t = f(X^t, F^{t-1:t+1}, c) \quad (1)$$

where $F^{t-1:t+1} = (F^{t-1}, F^t, F^{t+1})$ provides the model with the wind forcing at $t-1$, $t$, and $t+1$. Applied autoregressively, $f$generates continuous simulations of arbitrary length without reinitialization. The following sections describe the training data and preprocessing (a), the model architecture (b), the training procedure (c), and the validation strategy (d).

### *a. Data and preprocessing*

WaveGraph is trained on wave fields, atmospheric forcing, and static spatial features, all defined on the same unstructured mesh covering the Mediterranean Sea. The training wave variables are derived from the global dataset of Wadalkar et al. (2026), a bias-corrected version of the Mentaschi et al. (2023) simulation in which significant wave height fields were corrected against satellite observations, providing spectral wave fields at 3-hourly temporal resolution. The mesh features a spatial resolution ranging from approximately 2 km in coastal regions to a maximum of 50 km in the open ocean. The simulated variables are significant wave height (SWH), mean wave period (MWP), and mean wave direction (MWD). Atmospheric forcing consists of the 10 m zonal and meridional wind components from ERA5 (Hersbach et al., 2020), provided on a 0.25° regular grid and interpolated onto the unstructured mesh via bilinear interpolation at the same 3-hourly resolution. Each mesh node additionally carries static spatial features: bathymetry from GEBCO (GEBCO, 2023) and node spherical coordinates.

All dynamic variables are standardized using training-set statistics. MWD is decomposed into sine and cosine components before normalization, to remove discontinuities at the 0°/360° boundary. Bathymetry is scaled to [0,1], and node coordinates are encoded as trigonometric functions of latitude $\theta$ and longitude $\phi$, specifically $\cos(\theta)$, $\cos(\phi)$ and $\sin(\phi)$, projecting them onto the unit sphere to preserve spatial relationships without discontinuities. Input features at each node are then constructed by concatenating the dynamic variables over the temporal window with these static features.

*b. Model architecture*

The WaveGraph architecture is built around two graph representations of the spatial domain, following the multiscale framework proposed by Fortunato et al. (2022): a variable-resolution graph that preserves the variable resolution of the original unstructured mesh, and a uniform coarse graph that enables efficient modeling of basin-scale interactions. Information flows between the two resolutions through a pair of bipartite graphs, giving a total of four graphs that together define the multiscale processing pipeline. At each autoregressive step, node features are updated through successive message-passing stages on these graphs before being decoded into the predicted wave state.

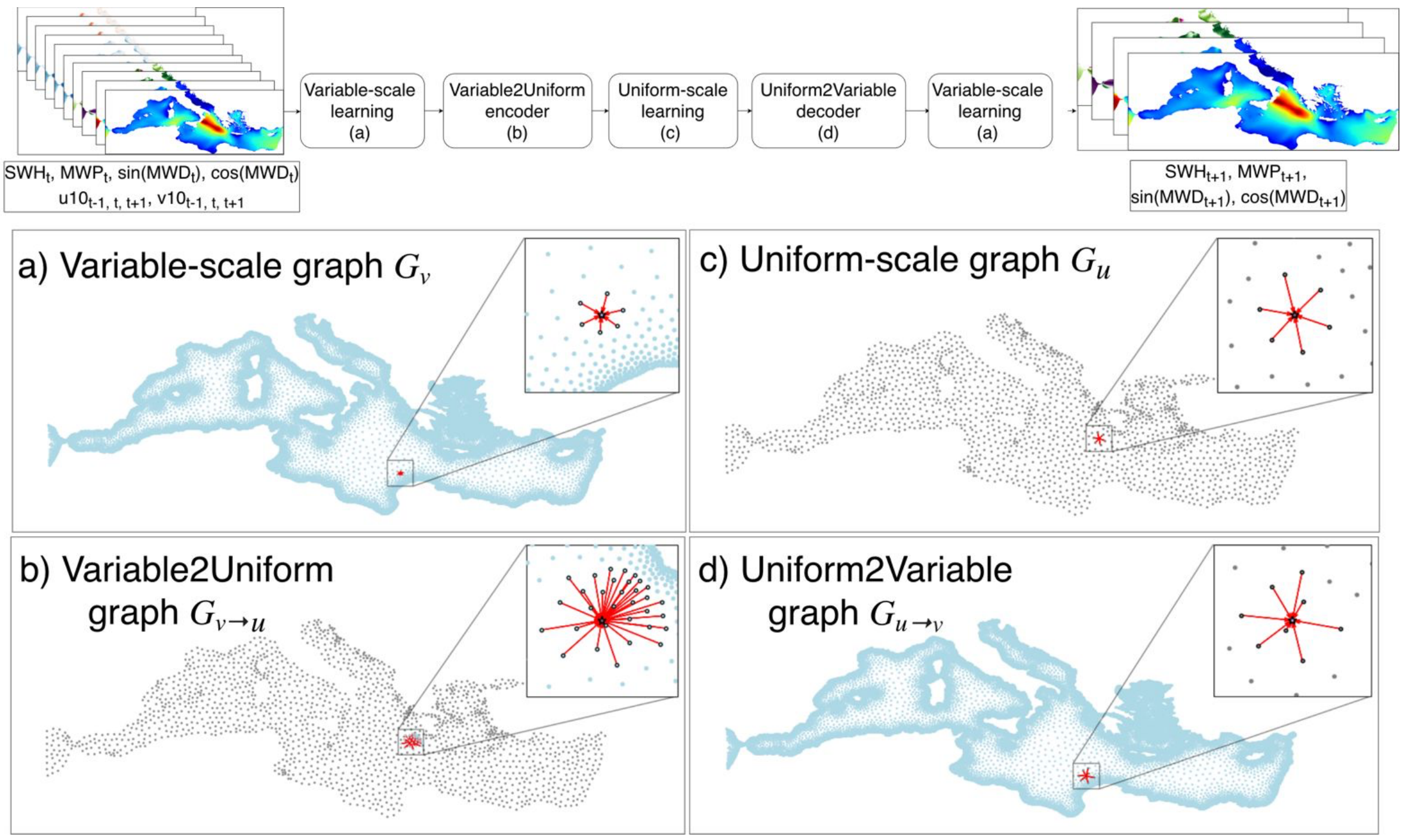

Fig. 1. Architecture of the WaveGraph encoder-processor-decoder pipeline. Input features (SWH, MWP, sin/cos(MWD) at t, and u10/v10 at t-1, t and t+1) on the variable-scale mesh are mapped through four learned stages, variable-scale learning (a), a variable-to-uniform encoder (b), uniform-scale learning (c), and a uniform-to-variable decoder (d), followed by a final variable-scale learning stage (a), to predict SWH, MWP, and sin/cos(MWD) at t+1.

### 1) Variable and uniform graph representations

The variable-resolution graph $G_v = (V_v, E_v)$ (Figure 1a) corresponds directly to the unstructured mesh of the wave model, where the set of nodes $V_v$ represents mesh points and edges $E_v$ follow the triangular connectivity. The uniform-resolution graph $G_u = (V_u, E_u)$ (Figure 1c) is defined on a triangular, quasi-uniform mesh at approximately 60 km characteristic resolution, providing a homogeneous representation of the domain that allows long-range correlations across the basin.

### 2) Inter-resolution connectivity

Exchange between the two representations is implemented through two bipartite graphs, $G_{v\to u}$ and $G_{u\to v}$ (Figures 1b and 1d). In the variable-to-uniform direction, each node of the uniform mesh aggregates information from its $k_{enc} = 32$ nearest neighbors from the variable mesh, based on Euclidean distance in 3D Cartesian coordinates. In the uniform-to-variable direction, each variable-mesh node receives information from its $k_{dec} = 8$ nearest uniform-mesh nodes. The asymmetry ($k_{enc} = 32$ versus $k_{dec} = 8$) reflects the difference in mesh density: aggregating from variable to uniform requires a broader stencil to adequately sample the denser fine mesh, while distributing from uniform to variable is a more local operation.

### 3) NODE AND EDGE FEATURES

Nodes in the variable-resolution graph carry dynamic wave variables, atmospheric forcing, and static features; nodes in the uniform graph carry geometric information only, acting as a computational substrate for large-scale message passing rather than as physical state carriers. Edge features encode the spatial relationship between connected nodes: the Euclidean distance and the components of the normalized displacement vector, computed in 3D Cartesian coordinates on the unit sphere and normalized by the maximum edge length across all graphs for consistent scaling.

### 4) MESSAGE PASSING

WaveGraph uses the Interaction Network formulation (Battaglia et al., 2016). For each edge connecting sender node $i$ to receiver node $j$, a message is computed as:

$$m_{ij} = \phi_e(h_i, h_j, e_{ij}) \tag{2}$$

where $h_i$, $h_j$ are the latent node representations and $e_{ij}$the edge features. Messages are aggregated by sum:

$$m_j = \sum_{i \in \mathcal{N}(j)} m_{ij}\,, \tag{3}$$

where $\mathcal{N}(j)$ denotes the set of sender nodes connected to node $j$. The node representation is then updated as

$$h_j' = \phi_v(h_j, m_j), \tag{4}$$

where $\phi_v$ is another learnable function. Both $\phi_e$ and $\phi_v$ are multilayer perceptrons (MLPs) with $N_h$ fully connected layers, Sigmoid Linear Unit (SiLU) activations, and layer normalization. The same latent dimension $d_h$ is used throughout, allowing for a consistent representational capacity across all stages of the pipeline.

### 5) INPUT EMBEDDING, OUTPUT MAPPING, AND PHYSICAL CONSTRAINTS

Node and edge features are first projected into the latent space of dimension $d_h$via dedicated embedding MLPs. After propagation through the message-passing stages, final node representations are mapped to the predicted wave increment through an output MLP. For variables that are strictly non-negative, SWH and MWP, a Softplus activation (Dugas et al., 2000) is applied at the output to enforce physical positivity, a constraint that was found to improve both training convergence and long-term autoregressive stability.

6) PROCESSING PIPELINE

The complete pipeline (Figure 1) proceeds as follows at each time step: $(i)$ $N_v$ message-passing steps on $G_v$ capture local coastal dynamics; $(ii)$ a single encoding step on $G_{v\to u}$ projects representations onto the uniform mesh; $(iii)$ $N_u$ steps on $G_u$ propagate information at basin scale; $(iv)$ a single decoding step on $G_{u\to v}$ maps representations back to the variable mesh; $(v)$ a final $N_v$ steps on $G_v$ refine local representations before output. This sequence allows WaveGraph to simultaneously resolve coastal complexity and large-scale wave propagation within a unified, computationally efficient framework. The values of all hyperparameters ($d_h$, $N_v$, $N_u$, $k_{enc}$, $k_{dec}$, and the number $N_h$ of hidden layers in the MLPs) are summarized in Table 1.

| Number of hidden layers in MLPs, $N_h$ | 1 |
|---|---|
| Variable-scale learning message steps, $N_v$ | 2 |
| Uniform-scale learning message steps, $N_u$ | 8 |
| Neighbors in Fine2Uniform graph, $k_{enc}$ | 32 |
| Neighbors in Uniform2Fine graph, $k_{dec}$ | 8 |

Table 1. Chosen hyperparameters for WaveGraph model

*c. Training procedure*

Rather than predicting the absolute wave state, WaveGraph is trained to predict the temporal difference $\Delta X^{t+1} = X^{t+1} - X^{t}$. This formulation focuses on learning on the local evolution of the system, which is typically smoother and of smaller magnitude than the full state, a choice that reduces the dynamic range the model must represent and facilitates stable autoregressive rollout (Lam et al., 2023; Holmberg et al., 2025).

Each temporal difference is standardized using training-set statistics before being passed to the loss:

$$\widetilde{\Delta X}^{t+1} = \frac{\Delta X^{t+1} - \mu_{\Delta X}}{\sigma_{\Delta X}} \quad (5)$$

The predicted increment is then mapped back to physical space and added to the last known state:

$$\hat{X}^{t+1} = X^t + \widetilde{\Delta \hat{X}}^{t+1} \cdot \sigma_{\Delta X} + \mu_{\Delta X} \quad (6)$$

Training minimizes a weighted mean squared error, with each variable *v* normalized by the variance of its increment $\sigma^2_{\Delta X_v}$ to balance contributions across variables with different dynamic ranges:

$$\mathcal{L} = \frac{1}{N_{nodes}} \sum_{i=1}^{N_{nodes}} \frac{1}{N_v} \sum_{v=1}^{N_v} \frac{\left(\widehat{\Delta X}_{i,v}^{t+1} - \Delta X_{i,v}^{t+1}\right)^2}{\sigma^2_{\Delta X_v}} \quad (7)$$

Area weighting was not applied because it would systematically downweigh coastal nodes, which carry the highest mesh resolution and the most complex wave dynamics, that is the opposite of what is physically desirable. Grid points west of the Strait of Gibraltar (longitude < -5.5°) are excluded from the loss, as they lie outside the Mediterranean domain of interest.

We conducted a series of sensitivity tests, summarized in Appendix A, to choose hyperparameters such as latent space dimension, number of past wave steps, number of wind forcing steps, and training set size. The dataset is split into training (2000-2005), validation (2006), and test (2007) periods. Optimization uses AdamW algorithm (Loshchilov & Hutter, 2017) with $\beta_1 = 0.9$, $\beta_2 = 0.999$, and a weight decay of 0.01. Learning rate follows a OneCycle policy schedule (Smith & Topin, 2017) starting from an initial value of $10^{-5}$, increasing to a maximum of $10^{-4}$ during the first 10 epochs, then decaying following a cosine annealing schedule. Training is conducted on two NVIDIA A100 GPUs and requires approximately 34 hours, using PyTorch, PyTorch Geometric, and Hugging Face Accelerate. The model is trained for 200 epochs with a local batch size of 2 and contains approximately 7 million parameters (Table A1). Both training loss and validation loss converged (Figure 2).

To assess the relative importance of wave history versus atmospheric forcing, we also performed an ablation experiment and trained a variant, named WaveGraph-wind-only, in which the previous wave state is removed from the input. This model relies solely on wind fields at $t-1$, $t$, and $t+1$ together with static spatial features, and serves as a controlled baseline to isolate the contribution of wave memory to model skill and long-term stability.

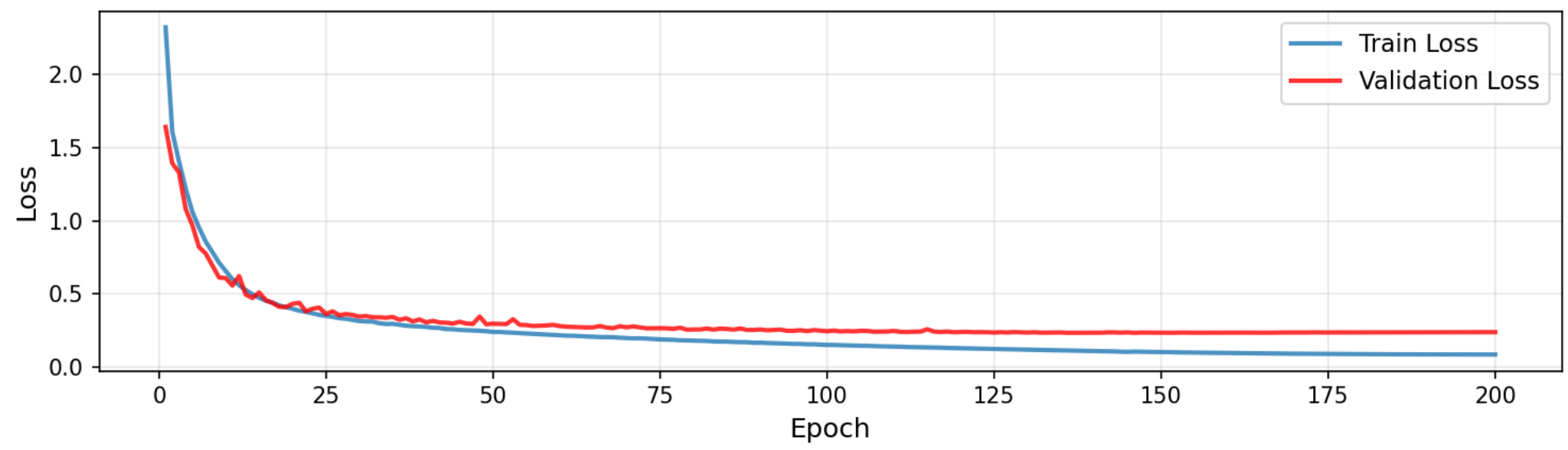

Fig. 2. Train and validation learning curves (full-configuration WaveGraph model)

*d. Model performance evaluation*

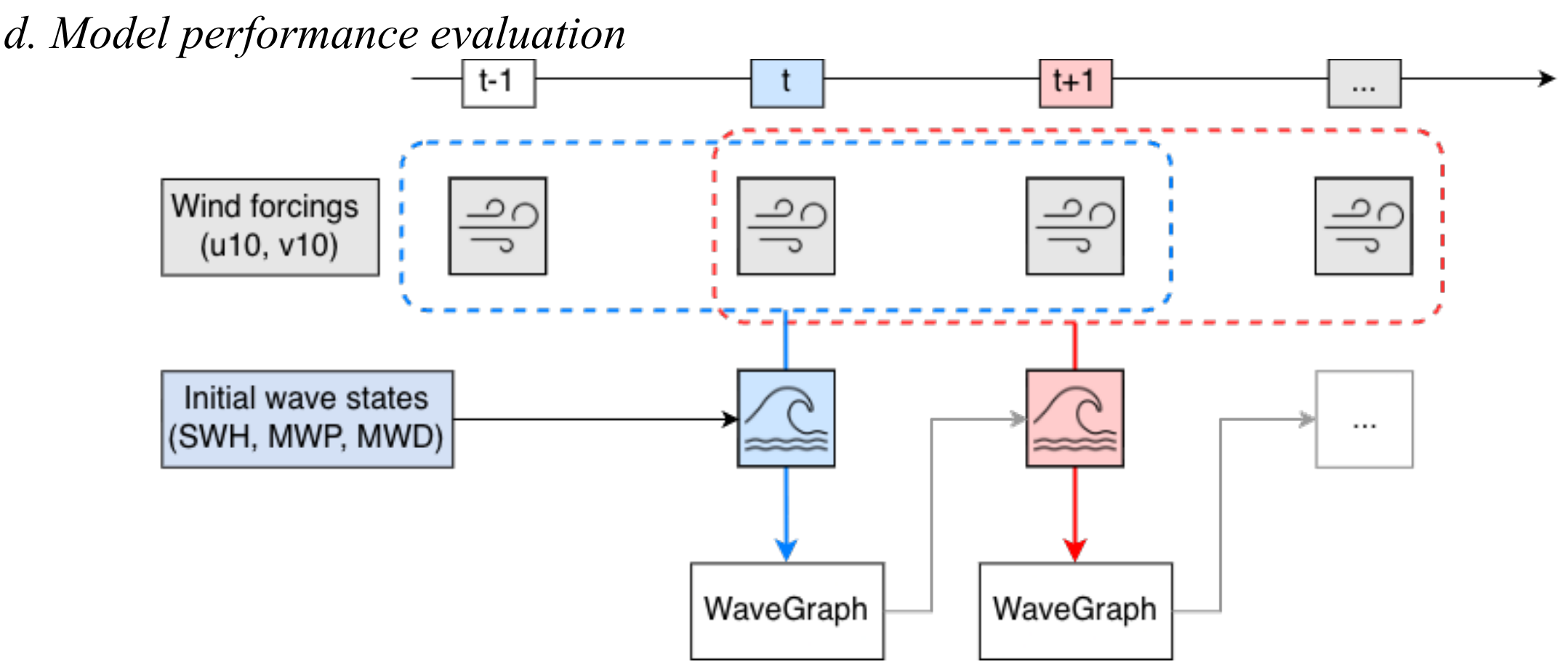

Fig. 3. Data workflow in autoregressive mode (full-configuration WaveGraph model)

Model performance is evaluated in autoregressive mode, where predicted wave states are recursively fed back as inputs while atmospheric forcing is prescribed from ERA5 (Fig. 3). This setup is consistent with the training objective and follows the evaluation protocol adopted in comparable data-driven wave and atmospheric models (Lam et al., 2023; Holmberg et al., 2025; Wang et al., 2025).

Over the 2007 test period, the model is initialized daily at 00:00 and 12:00 and run autoregressively for 960 hours (corresponding to 320 steps), allowing systematic assessment of error growth as a function of lead time on subseasonal scales. The lead time is defined as the number of autoregressive steps after initialization, corresponding to multiples of the 3-hour model time step.

Given the high degree of stability observed in these experiments, we further assessed whether WaveGraph could sustain accurate predictions over much longer time horizons. To this end, a continuous simulation was performed from 29 December 2006 to 31 December 2023 (a 17-year reconstruction at 3-hourly resolution) without any reinitialization. The initial wave state was set to zero and spun up over 16 time steps (48 hours) before evaluation began.

This experiment, which takes approximately 40 minutes on a single NVIDIA A100 GPU, tests whether the learned dynamics remain stable and physically consistent on decadal scales.

For the long-term simulation, model outputs are evaluated against independent buoy and satellite altimeter observations. Buoy data were obtained from the Global Ocean Delayed Mode Wave Product (Copernicus Marine In Situ TAC, 2025), complemented by records from the ISPRA Italian National Wave Network (https://www.mareografico.it/) and the Nausicaa station near Ravenna, Italy (https://simc.arpae.it/dext3r/). Quality-control flags provided with each product were applied, and the final set of stations is shown in Figure S1 (supplementary information). Satellite observations were obtained from the Global Ocean L3 Significant Wave Height product (European Union-Copernicus Marine Service, 2021), which provides homogenized SWH estimates with intercalibration corrections. Satellite tracks were processed individually, with observations failing quality control removed, land-contaminated measurements masked, and observation times rounded to the nearest hour. Model fields were interpolated to observation locations using barycentric interpolation on the unstructured mesh, followed by linear interpolation in time. SWH values smaller than 0.1 m for buoys and 0.5 m for satellite were excluded from the error evaluation.

*e. Evaluation metrics*

Model performance was evaluated using standard statistical metrics. Let $y_i$ and $\hat{y}_i$ denote observed and predicted values, respectively, with sample means $\bar{y}$ and $\bar{\hat{y}}$, and $N$ the number of observations. The following metrics were used:

- Root Mean Squared Error

$$\mathrm{RMSE} = \sqrt{\frac{1}{N}\sum_{i=1}^{N}(\hat{y}_i - y_i)^2} \tag{8}$$

- Mean Absolute Error

$$\mathrm{MAE} = \frac{1}{N}\sum_{i=1}^{N}|\hat{y}_i - y_i| \tag{9}$$

- Bias

$$\mathrm{Bias} = \frac{1}{N}\sum_{i=1}^{N}(\hat{y}_i - y_i) \tag{10}$$

- Corrected Mean Absolute Deviation (Campos-Caba et al., 2024)

$$\mathrm{MADc} = \mathrm{MAE} + \frac{1}{N}\sum_{i=1}^{N}\sum_{p}|\hat{y}_{i,p} - y_{i,p}| \tag{11}$$

where $p$ represents the percentiles of the distribution considered from 0 % to 100 % every 1 %. This additional term includes errors across different points of the distributions, making the metric more sensitive to extreme values.

- Pearson correlation

$$\mathrm{r} = \frac{\sum_{i=1}^{N}(\hat{y}_i - \bar{\hat{y}}_i)(y_i - \bar{y}_i)}{\sqrt{\sum_{i=1}^{N}\left(\hat{y}_i - \bar{\hat{y}}_i\right)^2 (y_i - \bar{y}_i)^2}} \tag{12}$$

- Scatter index

$$\mathrm{SI} = \sqrt{\frac{\sum_{i=1}^{N}\left[\left(\hat{y}_i - \bar{\hat{y}}_i\right) - (y_i - \bar{y}_i)\right]^2}{\sum_{i=1}^{N} y_i^2}} \tag{13}$$

- Normalized RMSE

$$\mathrm{NRMSE} = \sqrt{\frac{\sum_{i=1}^{N}(\hat{y}_i - y_i)^2}{\sum_{i=1}^{N} y_i^2}} * 100 \tag{14}$$

- To contextualise errors relative to natural variability, we compute the ratio RMSE/σ, where σ is the standard deviation of the observed wave field.

$$\sigma = \sqrt{\frac{1}{N}\sum_{i=1}^{N}(y_i - \bar{y}_i)^2} \tag{15}$$

Values below 1 indicate that model errors remain within the range of natural variability. RMSE, Pearson correlation and scatter index are also computed with respect to Wadalkar et al. (2026) as a function of lead time. For each lead time, predictions are compared with the reference model across all spatial nodes and all initializations, and the metrics are computed over this aggregated set of samples. We further evaluate the spatial distribution of RMSE and RMSE/σ across the model domain. WaveGraph skill against observations (buoys and satellite observations) is assessed through RMSE, MAE, Bias, MADc, Pearson correlation, SI and NRMSE.

## 3. Results

*a. Skills on subseasonal scales*

We first evaluate the performance of 960-hours reconstructions against the reference dataset (Wadalkar et al., 2026) following the experimental setup and lead time error computation described in Section 2d and 2e.

Figures 4a-c show the evolution of autoregressive error metrics for SWH as a function of lead time during the 2007 evaluation period. The error increases rapidly during the first steps, then the growth slows down, a behavior also observed for MWP and MWD (Figures 5 and 6). All metrics display a small-amplitude oscillatory signal with a 12-hour period, amounting to approximately 1% of the metric magnitude. After 960 hours of autoregressive reconstruction, the scatter index reaches approximately 0.06 for both SWH (~6 cm RMSE) and for MWP (~0.25 s RMSE), and 10° for MWD RMSE.  Pearson correlation remains high throughout the horizon, decreasing by only 0.004 for SWH and 0.2 for MWP after 960 hours.

The spatial distribution of the error at the final lead time (960 hours) is shown in Figures 4d-e, 5d-e, and 6b, and remains confined to specific regions of the domain. For SWH, the largest errors are mostly confined to coastal regions and around archipelagos, particularly the Aegean Sea, the Galite Islands (north of Algeria), and the Egadi Islands (west of Sicily), where RMSE locally reaches ~25 cm. For MWP, the spatial distribution of the error differs, with higher values observed near the domain boundary around the Strait of Gibraltar, the Egyptian coast near the Nile delta, the northern Ionian Sea, and the northern Adriatic Sea. For MWD, the error patterns are more spatially homogeneous but remain moderate across the basin, though greater errors are again found around the Strait of Gibraltar. Nevertheless, RMSE/σ generally remains less than 15% for SWH (Figure 4e) and less than 30% for MWP (Figure 5e) throughout most of the domain.

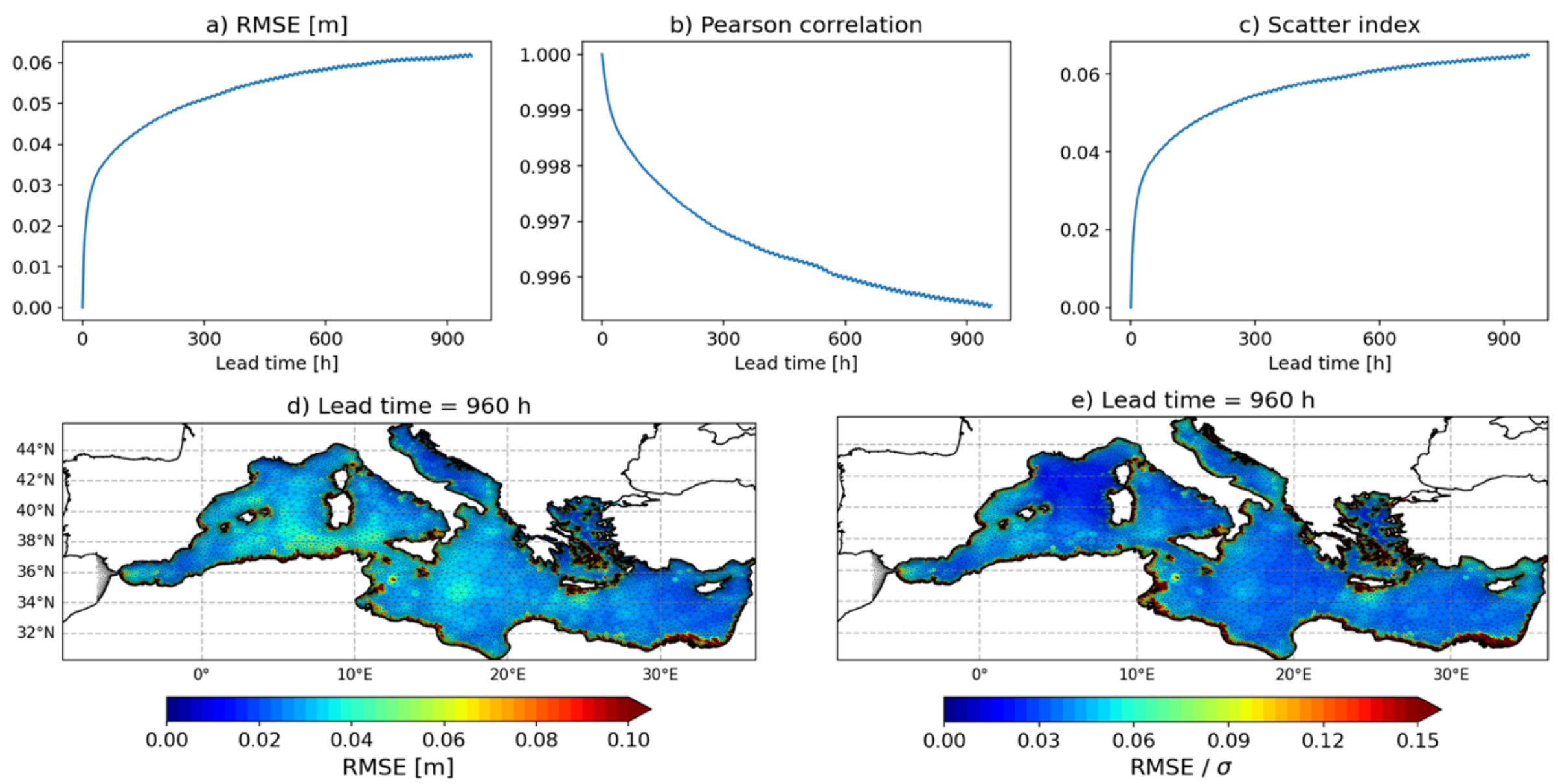

Fig. 4. Autoregressive error metrics for SWH against Wadalkar et al. (2026) during the 2007 testing period. Panels (a-c) show domain-averaged RMSE, Pearson correlation, and scatter index as a function of lead time, respectively. Panels (d) and (e) show the spatial distribution of RMSE and RMSE/σ at the 960 h lead time, computed at each mesh node.

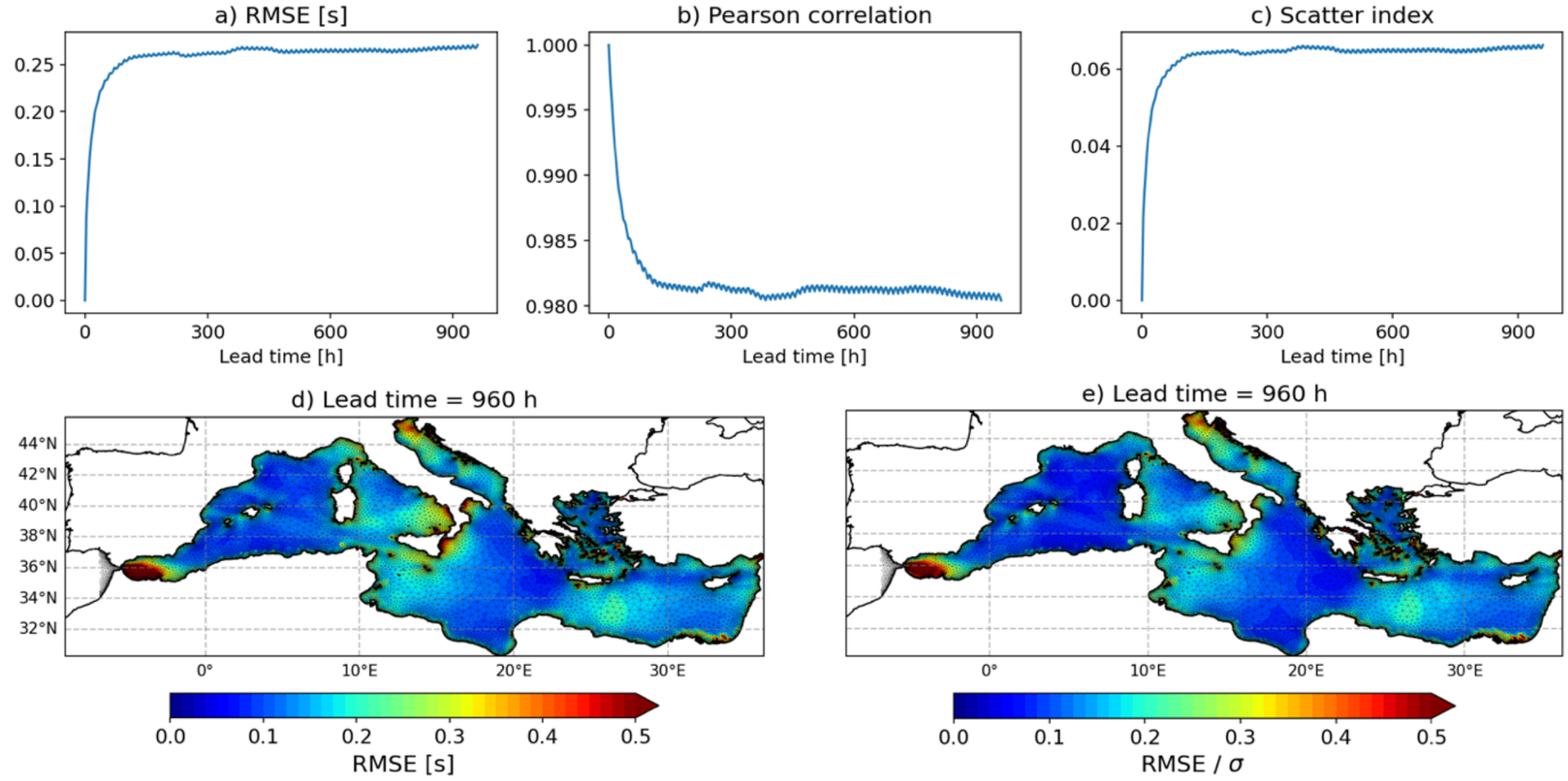

Fig. 5. Autoregressive error metrics for MWP against Wadalkar et al. (2026) during the 2007 testing period.

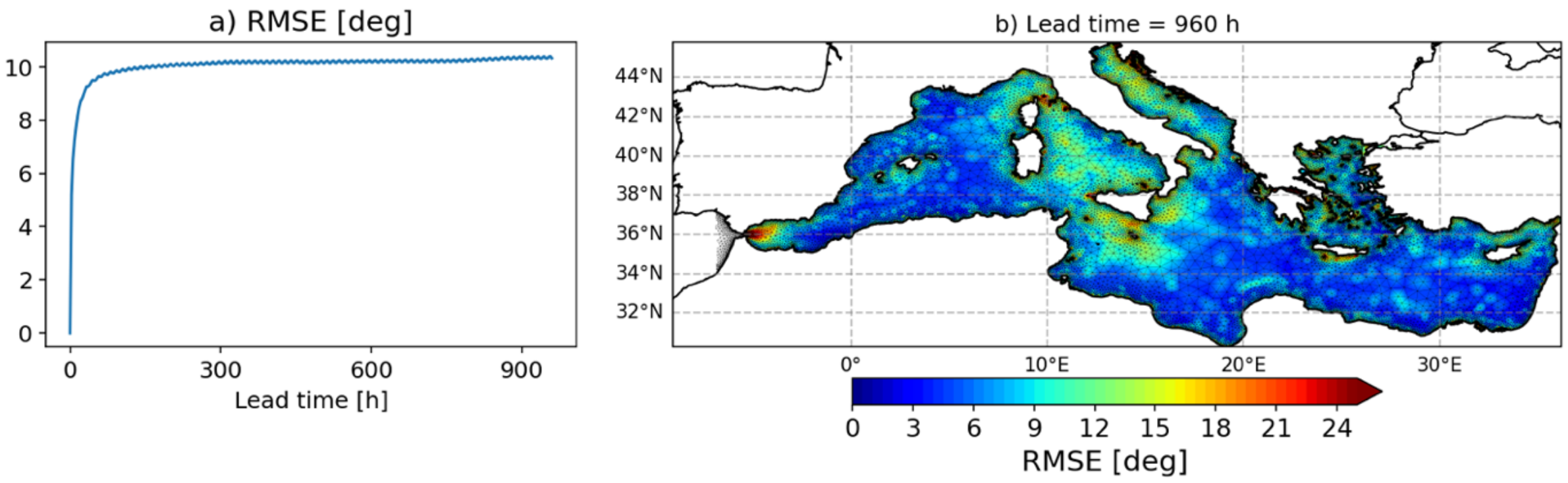

Fig.6. Autoregressive error metrics for MWD against Wadalkar et al. (2026) during the 2007 testing period.

*b. Decadal time scale reconstruction*

After assessing autoregressive performance over a 960-hours horizon, we extend the test set to evaluate the ability of WaveGraph to carry out a continuous long simulation spanning 17 years (2007-2023) without reinitialization. This experiment produces a three-hourly time series of wave fields over the full period, which is evaluated against the reference model (Wadalkar et al., 2026), wave buoys (Figure 7; Figure S5 for geographical reference) and satellite observations (Figure 8). Across both datasets, WaveGraph maintains a level of accuracy comparable to Wadalkar et al. (2026), with only minor differences in error metrics (e.g., ~1 cm in MADc), showing that the learned dynamics remain robust over decadal time scales without drift or instability. The largest errors, up to ~50 cm, are localized at a small number of stations (e.g., Palermo, Catania, and Capdepera), suggesting that discrepancies are mostly present in complex-bathymetry regions.

Satellite validation reveals spatial patterns consistent with the buoy analysis. As shown in Fig. 8, the overall performance metrics are comparable to those obtained from buoy observations. Higher errors are mainly concentrated in coastal regions and around islands, as well as in a region along the northern Algerian coast. This spatial consistency shows that the model generalizes beyond isolated observation points and captures basin-scale wave dynamics over decadal time scales with accuracy comparable to the reference model.

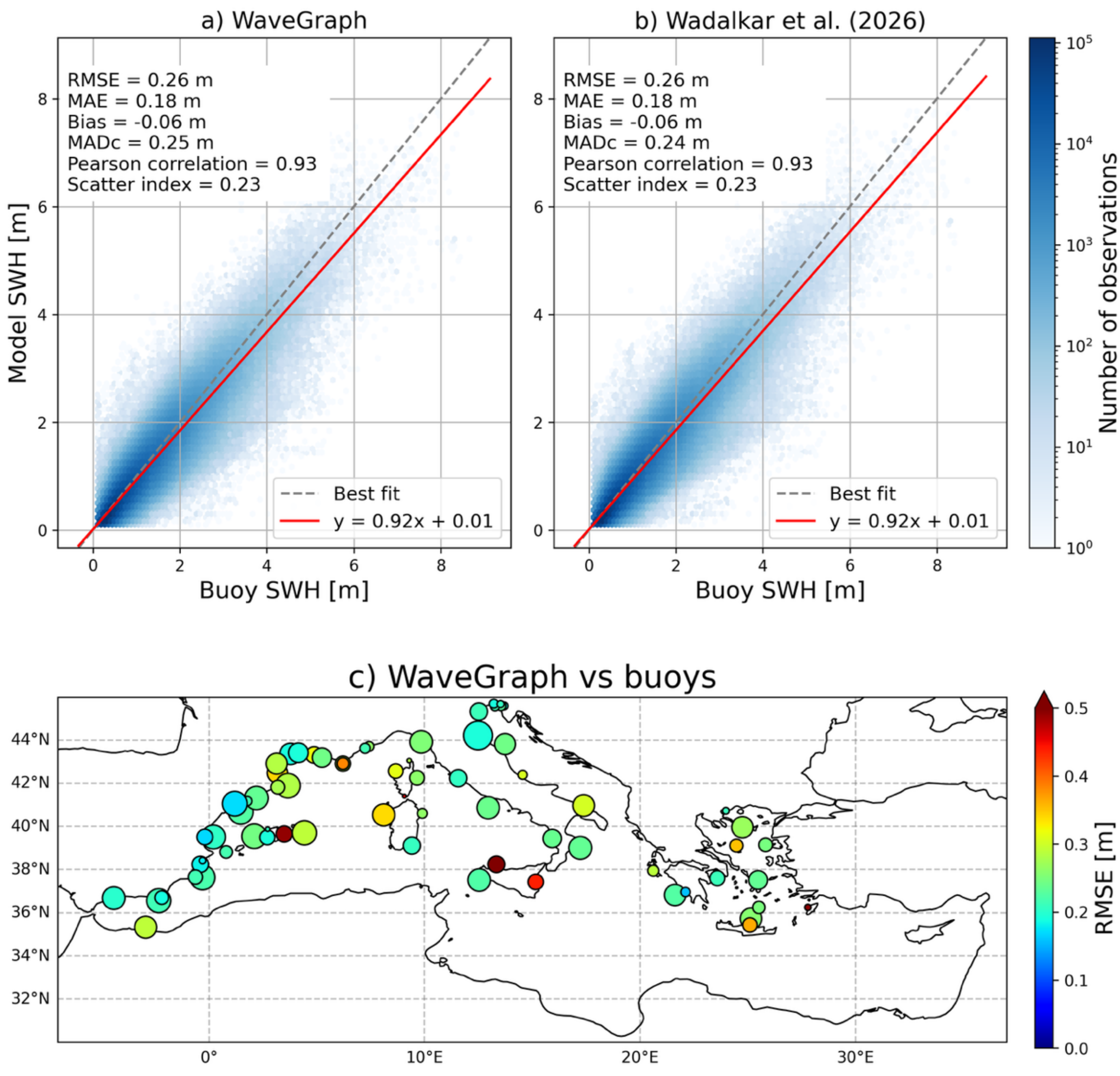

Fig. 7. Validation of SWH against buoys for the period 2007-2023 (N = 4331892 data points). The size of the dots in each location corresponds to the relative size of data of that buoy.

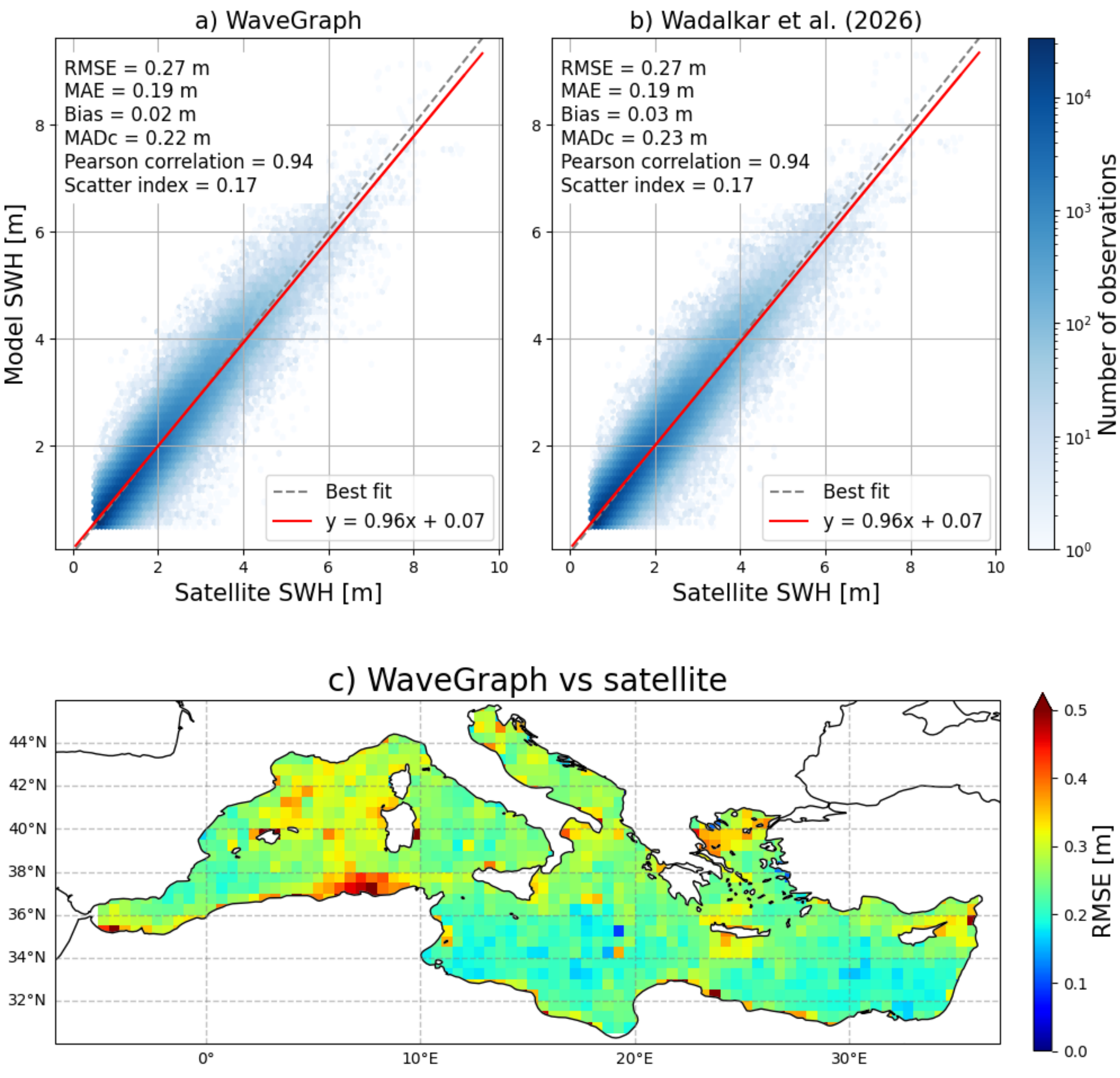

Fig. 8. Validation of SWH against satellite altimeter data for the period 2007-2023 (N = 1601479 data points).

*c. WaveGraph-wind-only ablation experiment*

To better understand the origin of the model stability, we performed an ablation experiment in which the wave history was removed from the input and the GNN learned solely by winds in the training period. This configuration tests whether the model can maintain stable predictions without explicit information about the previous wave state.

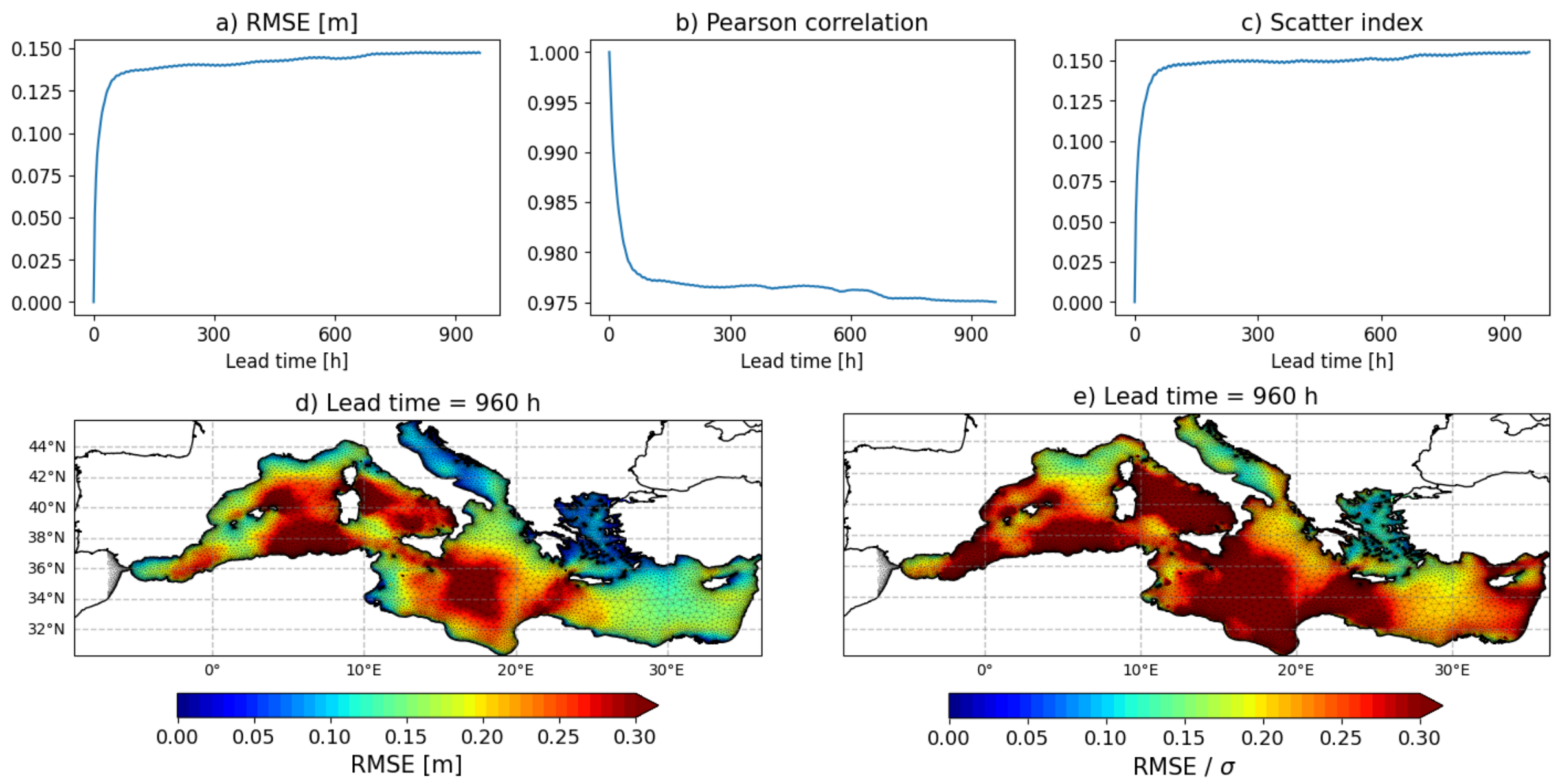

Fig. 9. Autoregressive error metrics for SWH against Wadalkar et al. (2026) during the 2007 testing period for the wind-only configuration.

Performance degrades compared to the full WaveGraph configuration; however, the model still produces physically consistent predictions and does not show drift or instability in autoregressive mode. Figures 9a-c show the evolution of SWH error metrics over the 2007 test period. The RMSE increases to ~15 cm and the scatter index reaches ~0.15, more than twice the error of the full configuration, while the correlation shows a larger but still limited degradation. All metrics reach a plateau and do not increase significantly at longer lead times, similarly to the full-configuration WaveGraph case.

The spatial distribution of the error highlights a strong dependence on the physical regime. The largest errors occur in open-sea regions, where RMSE exceeds 30 cm (Figure 9d) and reaches values of RMSE/σ > 0.3 (Figure 9e). In contrast, errors remain relatively low in wind-dominated basins such as the Adriatic Sea, the Aegean Sea, and the Gulf of Lion.

Validation against buoy and satellite observations (Fig. 10) confirms these patterns. The wind-only configuration shows a systematic degradation in skill, with buoy-based error metrics increasing by approximately 20% and RMSE reaching ~41 cm, corresponding to an increase of about 50% relative to the full-configuration model. The spatial distribution of normalized errors (Fig. 10c) further shows that performance degrades across most of the basin, with the exception of fetch-limited regions such as the Adriatic and Aegean Seas where skill is comparable and even slightly improves (e.g., ~2.5% less in Monopoli and Crete buoys).

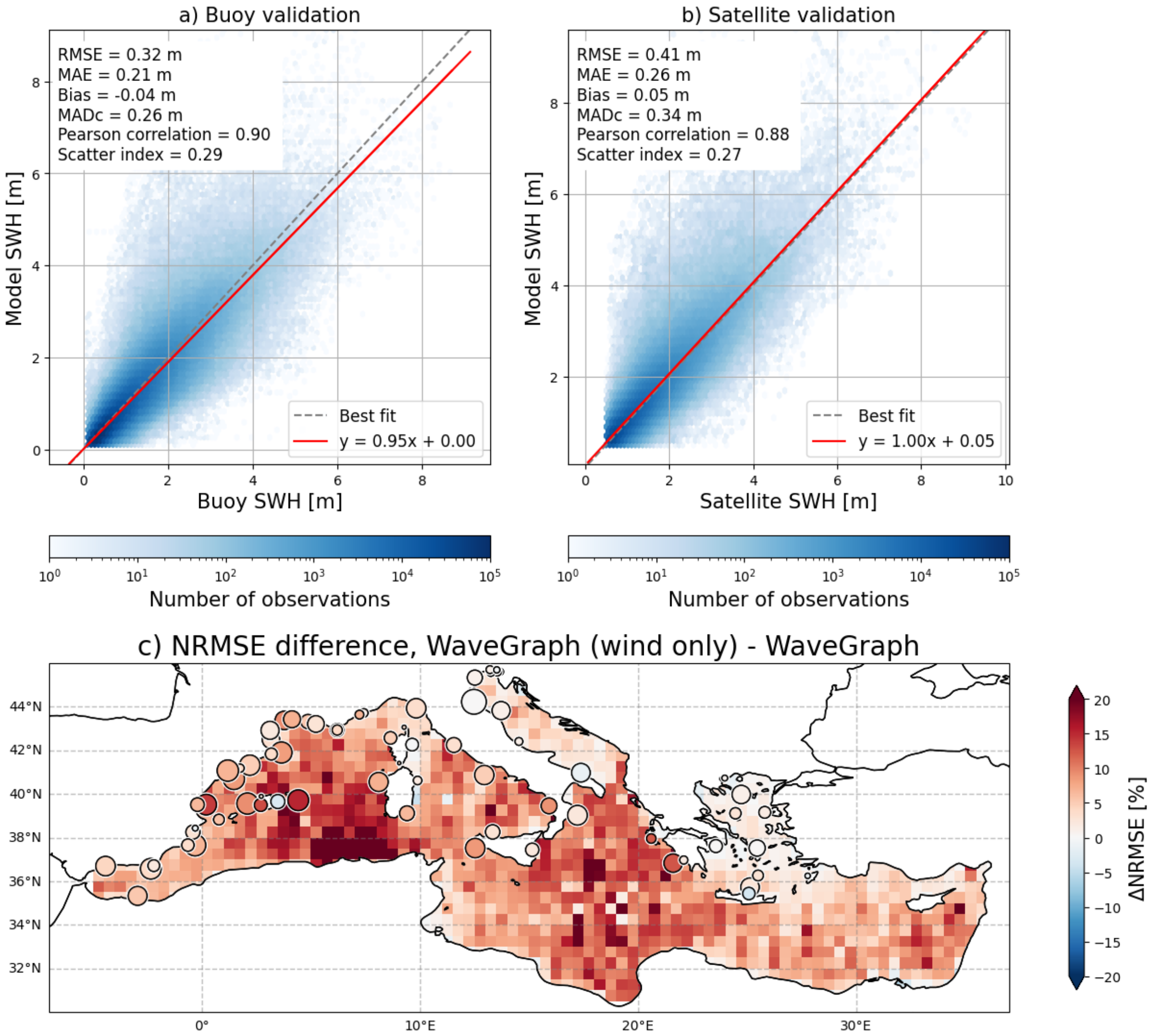

Fig. 10. Validation of SWH against buoy and satellite observations for the WaveGraph (wind only) model.

To clarify the local performance, we also analyzed time series at selected buoy locations, namely Nausicaa (Porto Garibaldi, Ravenna), Alghero (western Sardinia), and Dragonera (station 6100430; Balearic Islands). The selected stations represent different wave regimes within the Mediterranean basin, ranging from fetch-limited conditions in the northern Adriatic Sea to more energetic open-sea environments in the western Mediterranean.

Figure 11 shows that both WaveGraph and the WaveGraph-wind-only configuration are generally able to reproduce the temporal evolution of the observed wave field with good accuracy. WaveGraph tends to better reproduce the overall variability and the magnitude of the strongest events, particularly at Alghero and Dragonera, where swell and basin-scale propagation play a more important role. In these locations, the wind-only configuration overestimates the SWH peaks. Differences between the two configurations remain relatively small at Nausicaa, which is located in the fetch-limited Adriatic Sea.

These results show that wind forcing alone captures much of the local wave generation and dissipation, particularly in fetch-limited environments, but fails to reproduce swell and basin-scale propagation. This shows that wave history is important for WaveGraph to represent energy transport across the domain, beyond the immediate response to local wind forcing.

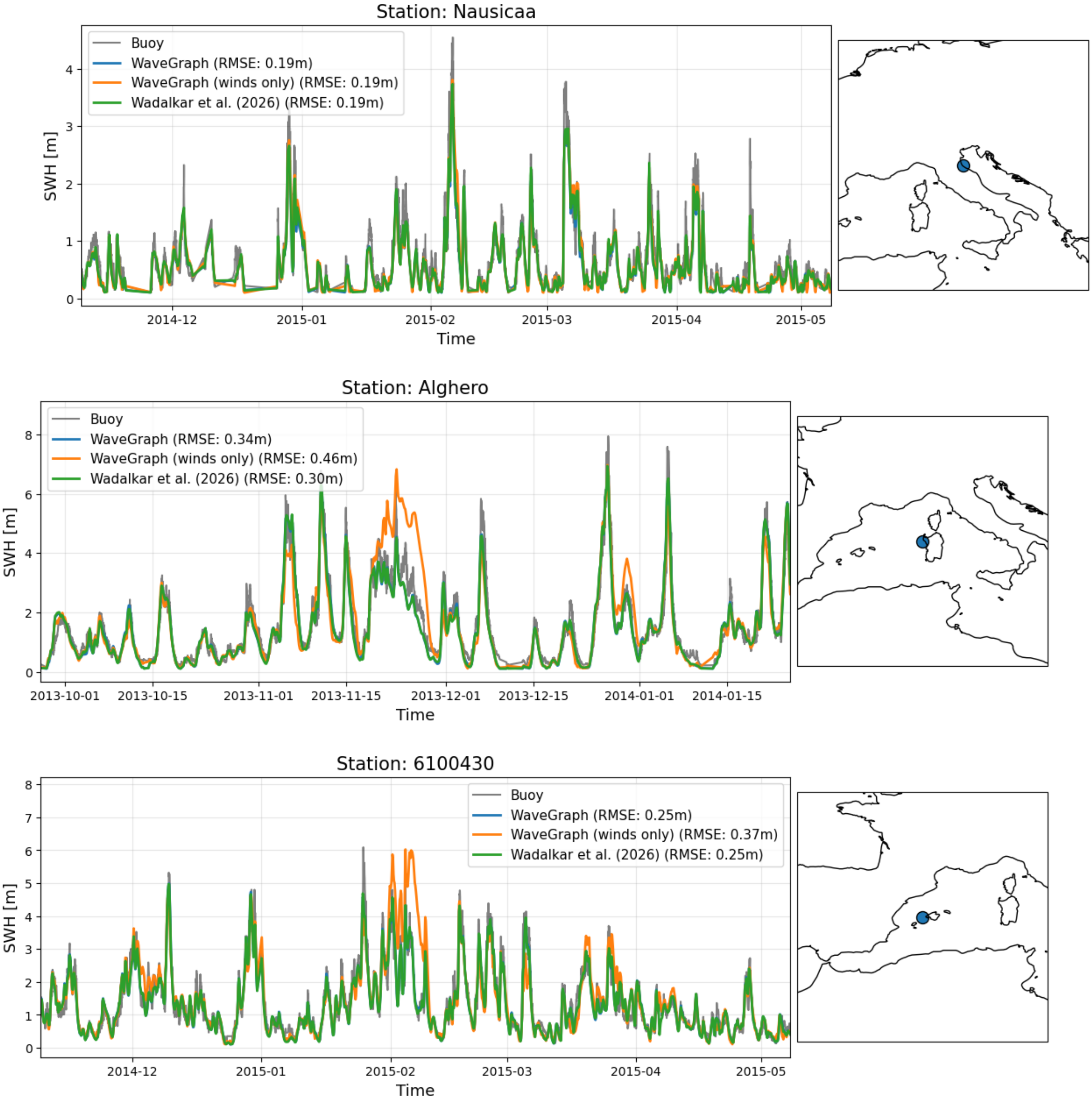

Fig. 11. Reconstruction and skill of the three models in three selected buoy stations for selected periods.

## 4. Discussion

This study presents, to our knowledge, the first application of a data-driven model of ocean wave dynamics operating directly on an unstructured mesh at basin scale and for decadal time scales. WaveGraph provides a computationally efficient surrogate of the Wadalkar et al. (2026) model while preserving its overall skill over decadal time horizons by using only a 6-years window for training (see Appendix A).

*a. WaveGraph stability*

The ability of WaveGraph to remain stable over long autoregressive rollouts follows from the intrinsic properties of wind-wave dynamics, which act on short timescales. The wave field responds rapidly to changes in atmospheric forcing and, in the absence of wind input, dissipates toward a resting state (Cavaleri et al., 2007). This short memory bounds error accumulation and allows stable autoregressive reconstructions over long horizons, in contrast to slower components of the Earth system.

Another important aspect of the WaveGraph model design is the enforcement of physical constraints, such as the positivity of wave variables (SWH and MWP; Section 2b). We found that this constraint plays an important role in improving stability during long autoregressive reconstructions. This behavior is consistent with recent work that applies similar constraints (e.g., Rectified Linear Unit, or ReLU, activations) in other AI-based ocean surface models (Hahner et al., 2026).

The ablation experiment provides an additional insight into the reasons of this stability. Results show that wind forcing alone provides a strong constraint on the temporal evolution of the wave field, allowing the model to maintain stable predictions even in the absence of explicit wave history as input variables during the training period. This is consistent with the physical role of wind as the primary source of energy input in wave dynamics. However, the inclusion of wave history remains important to accurately represent basin-scale wave propagation and nonlocal interactions, particularly in swell-dominated regions (Figures 4d, 9d, 10c, and 11). A similar role of wind in controlling autoregressive stability has also been observed in other data-driven models (Cao et al., 2023; Wang et al., 2025).

*b. Limitations*

Despite these promising results, some limitations remain. The largest errors occur in regions with complex, jagged coastlines, promontories, and small islands not resolved by the mesh. WaveGraph does not integrate a source term for unresolved obstacles (UOST), unlike the reference training model of Wadalkar et al. (2026), which applies the parameterization of Mentaschi et al. (2018) to account for islands and bathymetric features smaller than the computational grid. Without this information, the model likely struggles to represent energy propagation and dissipation in these areas, which explains at least part of the reduced accuracy observed there.

Additionally, a small-amplitude oscillation with a 12-hour period appears in the 2007 validation experiments (Figures 4-6 and 9). The same periodicity is also identified in the ERA5 wind forcing (Figure S2), which may be related to the 12-hour assimilation cycle of the 4D-Var system (ECMWF, 2026; Hersbach et al., 2020). A similar oscillatory pattern of the error, consistent with a 12-hour periodicity, can be observed in the lead-time error curves of other autoregressive models trained on ERA5 data (e.g., Oskarsson et al., 2024, Figures 16 and 17; McCabe et al., 2023).

## 5. Conclusion

In this work, we presented WaveGraph, a graph neural network designed to emulate basin-scale ocean wave dynamics on unstructured meshes. The model combines a multiscale graph representation with autoregressive reconstructions to reproduce the temporal evolution of multiple wave parameters over long time scales.

Results show that WaveGraph is able to reproduce the reference numerical model (Wadalkar et al., 2026) for 17 years with high accuracy while operating at a fraction of the computational cost (~2.5 minutes per year of reconstruction). Despite being trained only in a one-step forecasting setting, the model remains stable over very long autoregressive reconstructions, without showing visible drift or instability. Validation against buoy and satellite observations further shows that WaveGraph preserves the basin-scale skill of the reference model over decadal time scales.

The ablation experiments additionally show that wind forcing alone already constrains a large part of the wave evolution, particularly in fetch-limited environments, while the inclusion of wave history mainly improves the representation of basin-scale propagation and swell-driven dynamics.

Although some limitations remain, particularly in regions with complex coastal geometry, these results demonstrate that graph-based data-driven models can effectively emulate ocean wave dynamics over long time scales.

Several future developments are possible. Since the reference framework of Wadalkar et al. (2026) is global, extending WaveGraph to the global scale is a natural next step. The methodology is also directly applicable to other regional structured wave models and could be integrated into operational forecasting systems. Training on buoy and satellite wave observations alone is another option, given the relatively dense observing system coverage in the Mediterranean Sea.

More generally, the very low computational cost of the emulator opens the possibility for applications that are currently difficult with traditional spectral wave models, including large ensembles, probabilistic forecasting, long-term hazard mapping, and high-resolution climate-scale simulations. The high computational expense of spectral wave models has so far largely prevented their inclusion in coupled Earth system models (Brus et al., 2021). The emulator's low cost and fast response could change that, allowing wave-coupled physics to be reintroduced into Earth system models and extending the range of processes that fully coupled climate simulations can represent.

*Acknowledgments.*

The authors acknowledge funding by Horizon Europe Digital, Industry and Space (FOCCUS, grant no. 101133911) and the Research Council of Finland (grant no. 361902).

We acknowledge ISCRA for awarding this project access to the LEONARDO supercomputer, owned by the EuroHPC Joint Undertaking, hosted by CINECA (Italy).

*Data Availability Statement.*

Software used for preprocessing, training and inference is available at https://github.com/CMCC-Foundation/WaveGraph. Software used for validation against buoys and satellite altimetry is available at https://github.com/fedebenassi/scatter_waves. A copy of the Wadalkar et al. (2026) data, along with the mesh-interpolated ERA5 data, is available at https://doi.org/10.5281/zenodo.21526759 (Benassi et al., 2026).

## APPENDIX A

### Sensitivity tests

We assessed the sensitivity of the model to four design choices: latent space dimension (Figure A1), wave input history length (Figure A2), wind forcing temporal window (Figure A3), and training period length (Figure A4). All tests use the 2006 period for validation and report RMSE for significant wave height (SWH) and mean wave period (MWP), and mean angular error for mean wave direction (MWD), as a function of lead time.

We first tested three latent space dimensions (128, 256, 512), with wave input history fixed at one time step and wind forcing at t-1, t, t+1 (Figure A1). Increasing the latent space from 128 to 256 reduces SWH RMSE at 350 h from ~0.07 m to 0.05 m, MWP RMSE from 0.32 to 0.27 s, and MWD error from ~12° to ~10°. The further increase to 512 gives a smaller gain (SWH RMSE 0.04 m, MWP RMSE 0.25, MWD error 9°). Given the modest gain from 512 relative to the added computational cost, we used 256 in the final configuration.

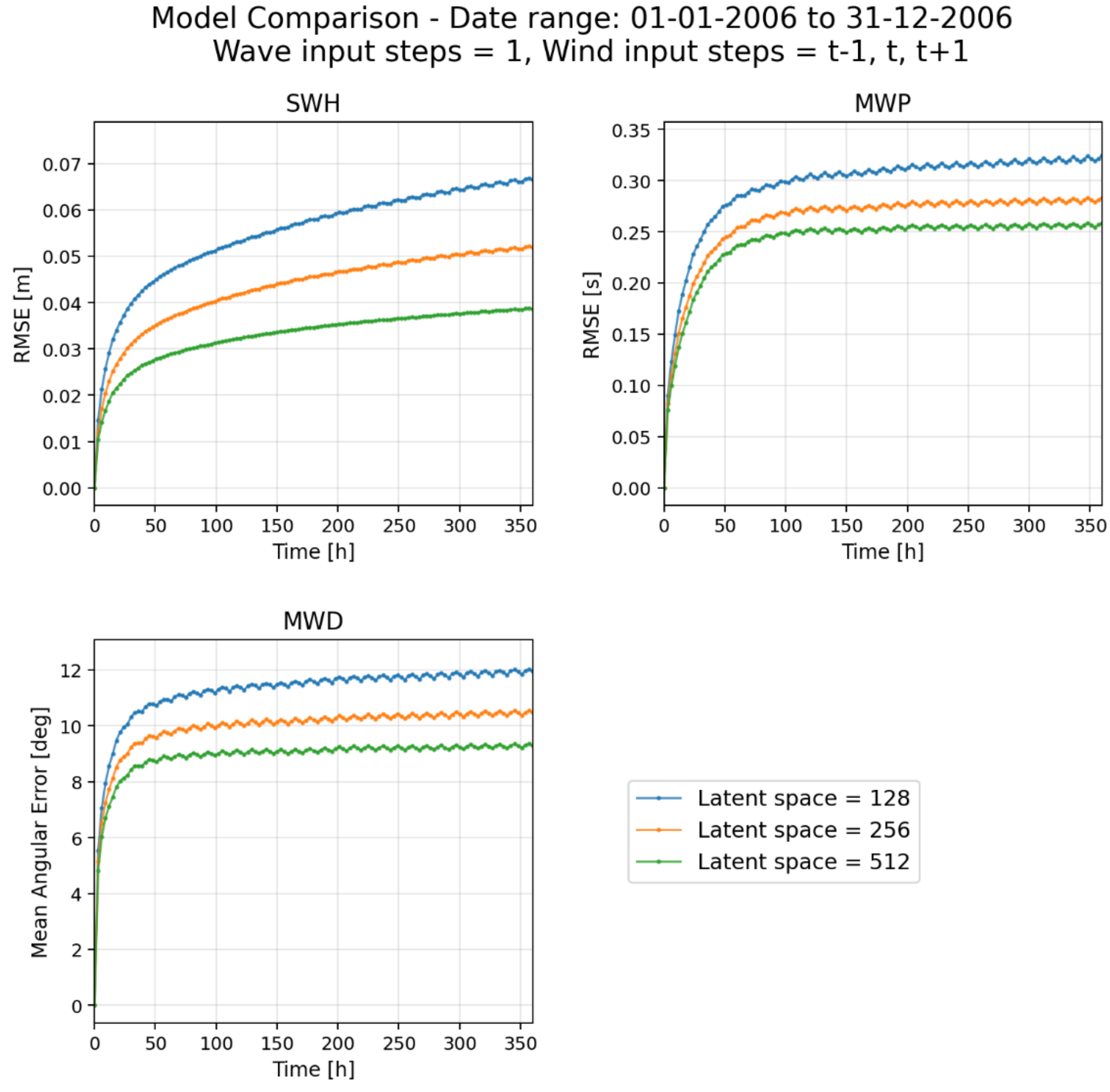

Fig. A1. Sensitivity test on the latent space dimension

With latent space fixed at 256, we then varied the wave input history from 1 to 8 time steps, and compared against a baseline with no wave input at all (Wavegraph-wind-only configuration, Figure A2). The wave-history configurations are nearly indistinguishable from each other across all three variables: SWH RMSE remains under 0.06 m, MWP under 0.5 s, and MWD error under 10°. Removing wave input changes this entirely: SWH RMSE saturates near 0.14 m within 50 hours, more than double the wave-history runs, and MWD error grows past 70° by 350 h against roughly 10° when wave history is present. MWP RMSE experiences the biggest growth, reaching 17.5 seconds RMSE. This shows that some wave memory is necessary for a stable forecast, but the length of that history has little effect beyond a single time step, which is what we used in the final model.

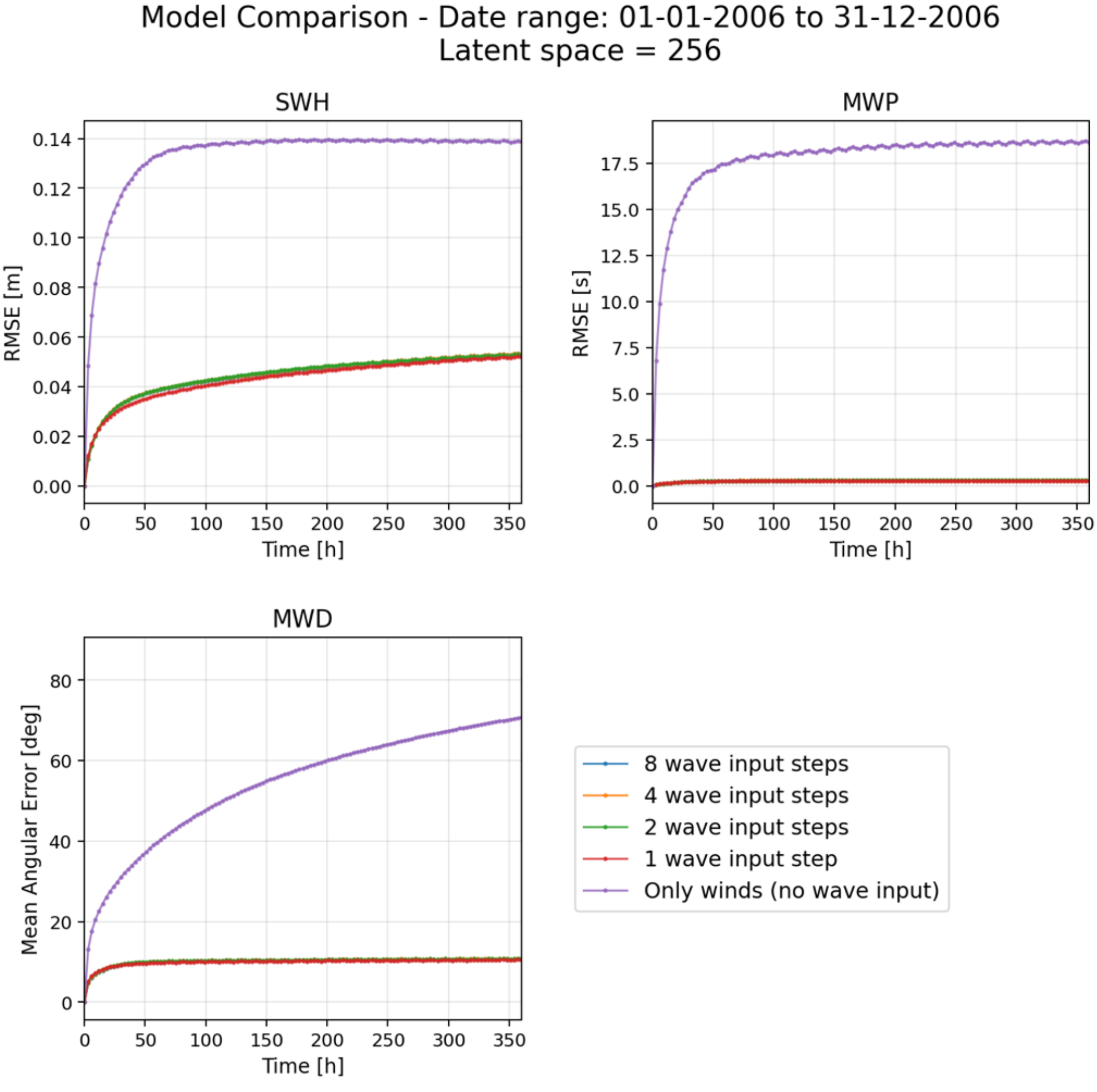

Fig. A2. Sensitivity test on the number of wave past steps

Next, we tested the wind forcing window, comparing t-1,t,t+1 against t,t+1 and t+1 only, with wave input fixed at one step (Figure A3). Including t-1 gives a small but consistent improvement across all three variables: at 350 h, SWH RMSE is 0.05 m for t-1,t,t+1 versus 0.065 m for t+1 only, and MWD error is 10° versus 12°. The gap between the three configurations is smaller than what we observed for the latent space and wave history tests, but it is present at every lead time, so we retained the full t-1,t,t+1 window in the final model.

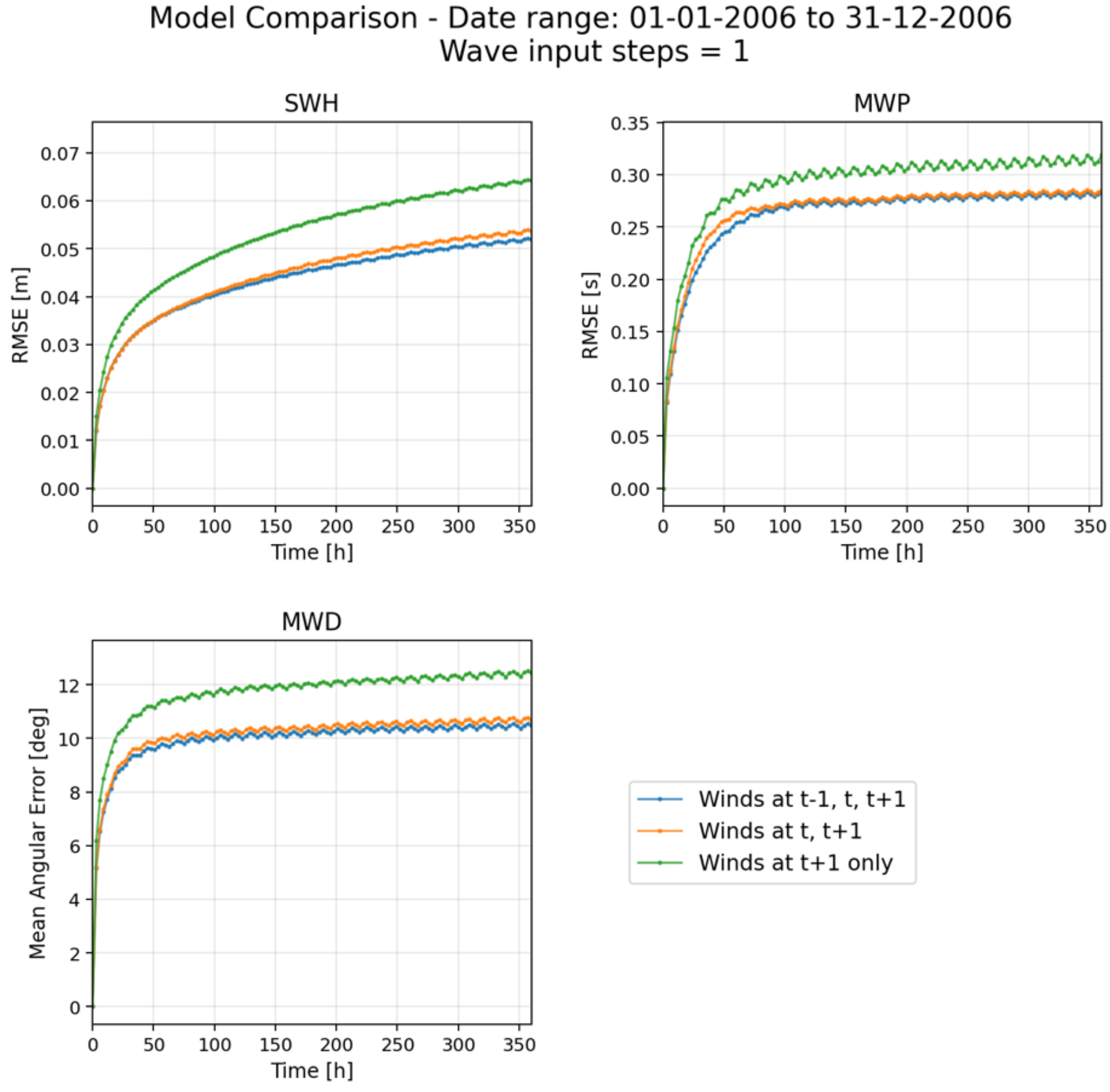

Fig. A3. Sensitivity test on the number of wind steps

Chosen the model hyperparameters and the input data, we tested the length of the training period, using 30 days (December 2005), one year (2005), six years (2000-2005), and eleven years (1995-2005), with latent space and wave input fixed at their selected values (Figure A4). Training on 30 days alone produces substantially worse forecasts, with SWH and MWP RMSE reaching 0.30 m and 1.0 s respectively and MWD error exceeding 30° by 350 h. Extending training to one year cuts these errors by roughly a factor of four. Further extending to six and eleven years gives comparatively small additional gains, with SWH RMSE around 0.04-0.05 m and MWD error around 9-10° for both. This suggests a plateau in the benefit of additional training data beyond a few years, and supports the use of the 6 years training period adopted in the final model.

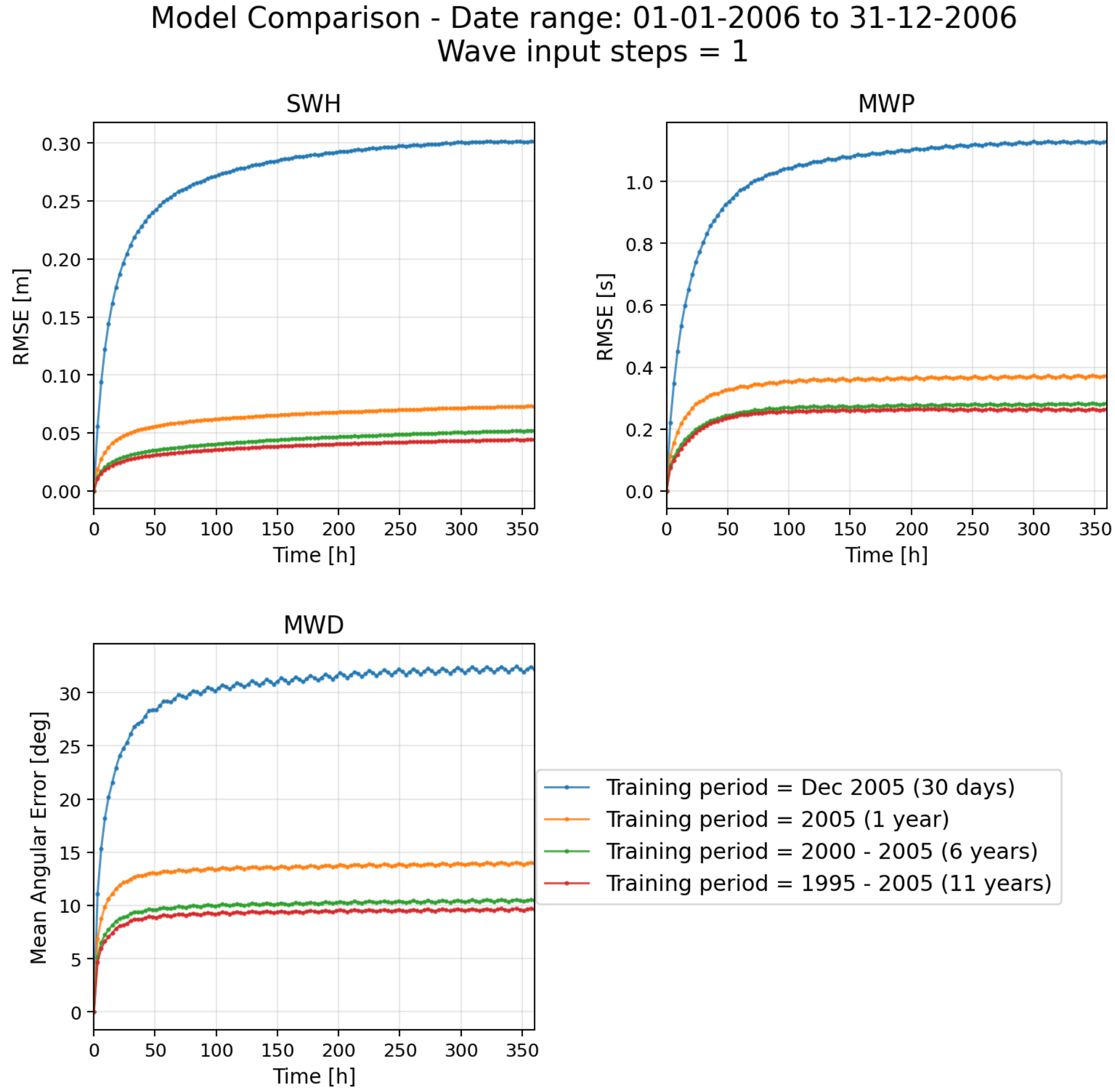

Fig. A4. Sensitivity test on the training set size

**Supplementary information for**

# Decadal wave reconstruction in the Mediterranean Sea with graph neural networks

Federica Benassi,[a,b] Lorenzo Mentaschi,[a,b] Salvatore Causio,[b] Daniel Holmberg,[c] Ivan Federico,[b] Nadia Pinardi[a,b].

[a] *CMCC Foundation – Euro-Mediterranean Center on Climate Change, Lecce, Italy*

[b] *Department of Physics and Astronomy, University of Bologna, Bologna, Italy*

[c] *Department of Computer Science, University of Helsinki, Helsinki, Finland*

*Corresponding author*: Federica Benassi, federica.benassi@cmcc.it

Content:

-

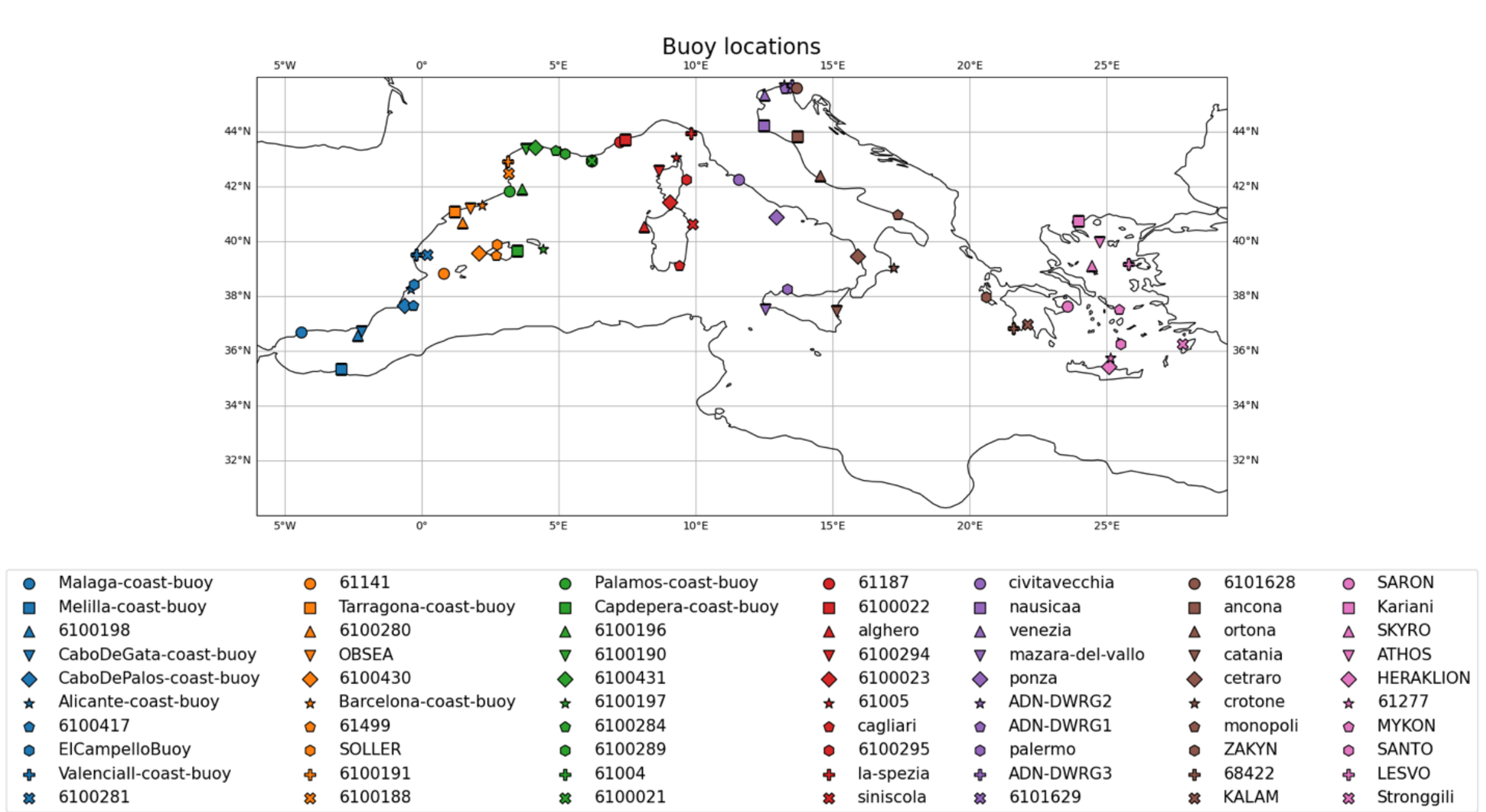

Fig. S1. Buoys used for validation

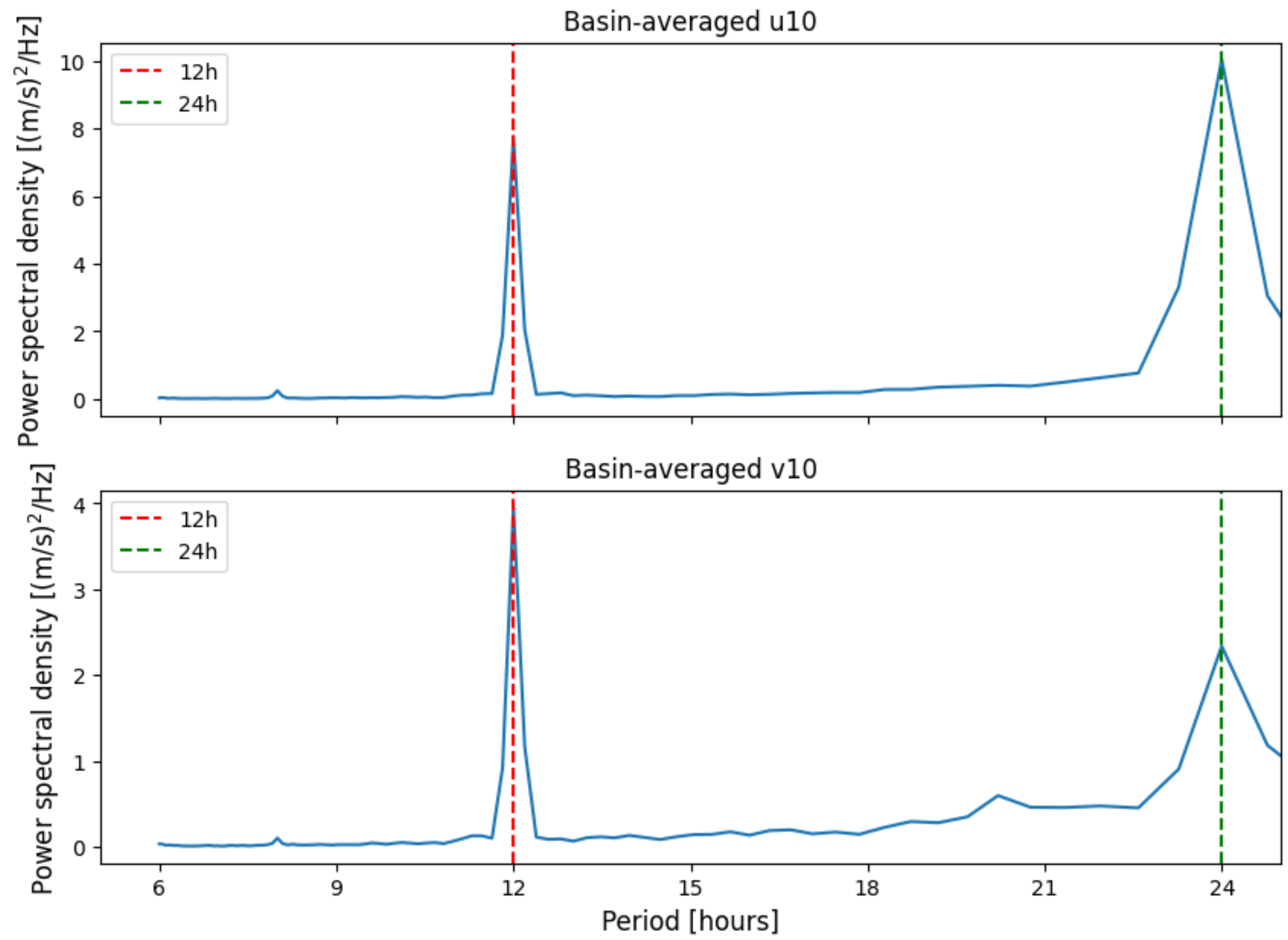

Fig. S2. Power spectral density (Welch's method) of the basin-averaged 10-m wind components, u10 (top) and v10 (bottom), as a function of period. Dashed lines mark the 12h (red) and 24h (green) periods.